\documentclass [12pt]{article}
\usepackage{amsmath}
\usepackage{amssymb}
\usepackage{graphics}
\usepackage{graphicx}

=240mm

\begin{document}
\title{Multi-energy radiography on the basis of ``scintillator-photodiode''
detectors}
\author{S.V. Naydenov, V.D. Ryzhikov, B.V. Grinyov, \\ E.K. Lisetskaya,
D.N. Kozin, A.D. Opolonin \\ {\it Institute for Single Crystals of
NAS of Ukraine}, \\ {\it 60 Lenin ave., 61001 Kharkov, Ukraine} }

\date{\empty}

\maketitle

\begin{abstract}
For reconstruction of the spatial structure and thicknesses of
complex objects and materials, it is proposed to use
multi-radiography with detection of X-ray or gamma-radiation by
combined detector arrays of scintillator-photodiode type.
Experimental studies have been carried out of the energy
dependence of sensitivity of dual-energy inspection systems based
on scintillators $ZnSe(Te)$ and $CsI(Tl)$. 
\end{abstract}

\section{Introduction}
The digital radiographic method is one of the main directions of
modern industrial non-destructive testing (NDT), e.g., \cite{1}.
Technical diagnostics (TD) are based on scanning (linear or
planar) and subsequent topography of the three-dimensional
structure of the object. It is often needed to carry out
quantitative analysis of the internal structure of materials. When
the geometry is complex, as well as for systems of variable
thickness, multi-layered, multiply connected or multi-component
structure, conventional NDT methods could be insufficient. The use
of more informative and more complex tomographic methods is not
always possible due to technical or economical reasons. Important
progress can be achieved here in relationship with a multi-energy
radiography (MER). Monitoring with separate detection of radiation
at several energies can give additional information on the
internal structure of the studied object. A block diagram of such
method is presented in fig.~1.
\begin{figure}[ht]
\begin{center}
\includegraphics*[scale=0.8]{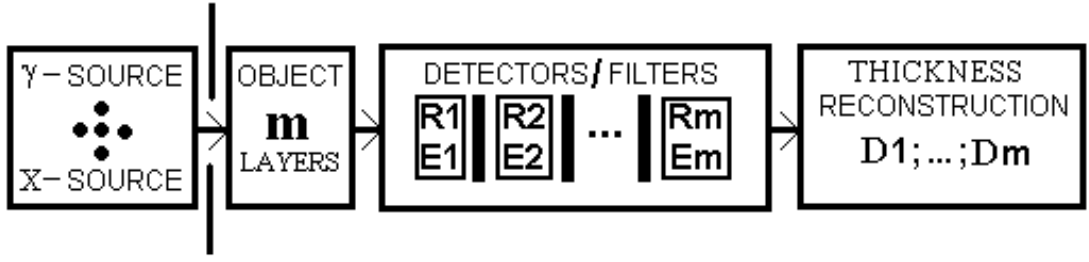}
\end{center}
\caption{Schematic diagram of multi-energy radiography of
thickness (defects).} \label{fig:Fig1}
\end{figure}

\section{MER Thickness Reconstruction}
In developing of the said aspect of MER, especially efficient are
simple schemes of two- and three-energy monitoring. It follows
from the theory that the number of reconstructed thicknesses is
the same as the multiplicity of radiography, i.e., the number of
separately detected radiation ranges. Consequent local scanning of
the object allows us to reconstruct the profile of its internal
three-dimensional structure also in the case of variable
cross-section of the components that form it. To determine
thickness of separate components or size of inclusions of each
specific material, one has to assume their effective atomic
number~$Z$ and density~$\rho $ to be approximately known. Or,
linear attenuation coefficients should be specified for
corresponding substances. For independent determination of
these~$Z$ and~$\rho $, it is also possible to use means of MER
\cite{2}. Theoretical model for thickness reconstruction by means
of MER uses the universal character of exponential attenuation of
the radiation in objects and detectors. Passing over to
logarithmic units of the detected signal normalized to the
background value (when the object is absent), radiography
equations can be presented in a simple form
\begin{equation}\label{eq1}
R_{i} = {\sum\limits_{j = 1}^{M} {\mu _{ij} D_{j}} } \;;\quad i =
1,\ldots ,m \;;\; j = 1,\ldots ,M  \quad ;
\end{equation}
where $ \mu _{ij} = \rho _{j} \left[ \tau \left( E_{i} \right)
Z_j^4 + \sigma \left( E_{i} \right) Z_{j} + \chi \left( E_{i}
\right) Z_j^2 \right] $ and $ R\left( E_{i} \right) \equiv R_{i} $
are reflexes (registration data) at corresponding maximum
absorption energies within each monitoring range. Unknown are
thicknesses $D_{i}$. Matrix $\mu _{ij} $ (of linear attenuation
coefficients) will be specified, with energy dependencies on
photo-effect $\tau $, Compton scattering $\sigma $ è and pair
generation effect $\chi $. In the medium energy range up to $0.5\,
MeV $, the latter scattering channel can be neglected. Solving the
linear system (\ref{eq1}) is the inverse problem of MER. To obtain
its unique solution the number of layers $m$ should correspond to
the order $M$ of multiplicity, $m = M$. The general solution has
the form
\begin{equation}\label{eq2}
D_{i} = \sum\limits_{j = 1}^{m} {\mu _{ij}^{-1} R_{j}}  \;;\quad
\det \mu _{ij} \neq 0 \quad ,
\end{equation}
where $\mu _{ij}^{-1} $ is the inverse matrix. In the case of
2-MER
\begin{equation}\label{eq3}
D_{1} = {\frac{{\mu _{22} R_{1} - \mu _{12} R_{2} }}{{\mu _{11}
\mu _{22} - \mu _{12} \mu _{21}} }}\;;\quad D_{2} = {\frac{{\mu
_{11} R_{2} - \mu _{21} R_{1}} }{{\mu _{11} \mu _{22} - \mu _{12}
\mu _{21}} }} \quad .
\end{equation}
In the general case, for determination of $D_{i} $ it is necessary
and sufficient that determinant $\det \mu _{ij} \ne 0$. This
implies a physical condition for MER feasibility:
\begin{equation}\label{eq4}
{\rm for \; all} \;\; i \neq j \quad \Longrightarrow \quad \left|
{E_{i} - E_{j}}  \right| \gg \delta E_{noise} \;,
\end{equation}
where $\delta E_{noise} $ is the total noise level in the system
expressed in energy units. For one-energy radiography, separation
of the reconstructed ``images'' of the complex object by scanning
at one camera angle is not possible. This experimentally and
theoretically proven fact corresponds to the uncertainty of
expressions (\ref{eq3}) when their denominator becomes zero at
$E_{1}=E_{2}$.
\section{Sensitivity of Multi-Radiography}
For practical developments of the MER, an important factor are
detector sensitivity, $S=dR\left/ dD\right.$, and relative
sensitivity, $T={\delta }\left/{D}\right.$, of the inspecting
system. Here $\delta$ is the minimum wire thickness that could be
detected on the main detection background (accounting for noises);
$D$ is the fixed thickness of the tested sample. As it follows
from Eqs.~(\ref{eq2})--(\ref{eq3})
\begin{equation}\label{eq5}
S \propto \left[K(E)\:\mu (E)\right] \;;\quad T \propto \frac{\rm
const}{K(E)\:\mu (E)\,D} \quad \Longrightarrow \quad S\sim T^{-1}
\;.
\end{equation}
where $K(E)$ -- conversion efficiency of the detectors. In the
Concern ``Institute for Single Crystals'', combined detectors of
``scintillator-photodiode'' type have been developed, which are
characterized by improved detector sensitivity. In assemblies of
2-radiography, tellurium-doped zinc selenide $ZnSe(Te)$ is used as
scintillator for the low-energy detector (LED). For the
high-energy detector (HED) scintillator $CsI(Tl)$ is used.
$ZnSe(Te)$ has lower effective atomic number, $Z_{L} = 32$, as
compared with cesium iodide, $Z_{H} = 54$, but its density is high
enough to ensure efficient absorption of the radiation in the low
energy region. This allows its use not only as a detector, but
also as an energy filter cutting off the low-energy part of the
spectrum. Efficiency of such filter with photo-effect is not less
than $ \propto \left( {1 - \left( Z_{L}\left/ Z_{H}\right.
\right)^{4}} \right)\approx 88\% $. Light output of $ZnSe(Te)$ can
reach $100-130\%$ with respect to $CsI(Tl)$ at absorbing thickness
of $0.1-1.0\,mm$. Optimum thickness values of scintillators for
the 2-MER with $U_{a}=140\,kV$ are: for LED $ZnSe(Te)$ --
$0.6\,mm$; for HED $CsI(Tl)$ -- $4\,mm$. As a result, all this
ensures substantial advantages of zinc selenide for radiation
detection in the $20-80\,keV$ range as compared with other
scintillators. Fig.~2 show results of our measurements of the
detecting sensitivity $S$ for with various combined
scintielectronic detector arrays. The data obtained confirm
advantages of the chosen design of the dual-energy inspection
system.
\begin{figure}[ht]
\begin{center}
\includegraphics*[scale=0.80]{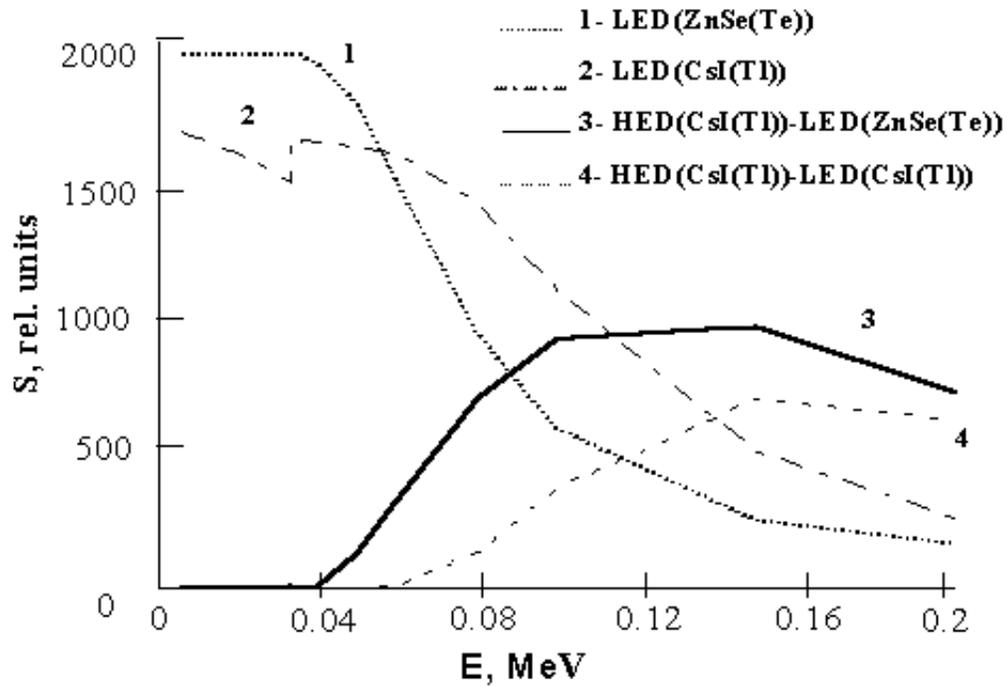}
\end{center}
\caption{Energy dependence of detector sensitivity of two-level
inspection systems. The X-ray source used had anode voltage
$U_{a}=40-180\,kV$ and current $I_{a}=0.4\,mA$.} \label{fig:Fig2}
\end{figure}
\section{Conclusions}
The developed scheme of MER can be directly used for different
control evaluations, especially in topography of several
surimposed ``layers''(or defects) or when analysis under different
camera angles is impossible. Quantitative determination of
thicknesses in a many-component structure makes it possible to
physically discern between physically surimposed parts of one and
the same piece of object. This substantially increases contrast
sensitivity of MER as compared with conventional methods, which is
important not only for technology, but also for medical
applications.

\end{document}